\documentclass{aa}
\usepackage{txfonts}
\usepackage{graphicx}
\usepackage{mathrsfs}
\usepackage[usenames]{color}
\usepackage{subfigure}

\newcommand{\cM}{\mathcal{M}}

\newcommand{\bfl}{\begin{flushleft}}
\newcommand{\efl}{\end{flushleft}}

\newcommand{\ud}{\mathrm{d}} 
\newcommand{\be}{\begin{equation}}
\newcommand{\ee}{\end{equation}}

\newcommand{\zhh}{$\mathrm{\zeta^{H_{2}}}$}

\newcommand{\Ezhh}{\mathrm{\zeta^{H_{2}}}}

\newcommand{\ea}{$\mathrm{\eta_{AD}}$}
\newcommand{\eh}{$\mathrm{\eta_{H}}$}
\newcommand{\eo}{$\mathrm{\eta_{O}}$}

\begin{document}

\title{The role of cosmic rays on magnetic field diffusion\\and the formation of protostellar discs}

\author{M. Padovani\inst{1,2,3}, D. Galli\inst{3}, P. Hennebelle\inst{4}, B. Commer\c con\inst{5}, and M. Joos\inst{4}}

\authorrunning{M. Padovani et al.}

%\titlerunning{TITLERUNNING}

\institute{Laboratoire Univers et Particules de Montpellier, UMR 5299 du CNRS, Universit\'e de Montpellier II, place E. Bataillon, cc072, 34095 Montpellier, France
\and
Laboratoire de Radioastronomie Millim\'etrique, UMR 8112 du CNRS, \'Ecole Normale Sup\'erieure et Observatoire de Paris, 24 rue Lhomond, 75231 Paris cedex 05, France\\
\email{Marco.Padovani@lupm.univ-montp2.fr}
\and
INAF--Osservatorio Astrofisico di Arcetri, Largo E. Fermi 5, 50125 
Firenze, Italy\\
\email{galli@arcetri.astro.it}
\and 
CEA, IRFU, SAp, Centre de Saclay, 91191 Gif-Sur-Yvette, France\\
\email{[hennebelle,marc.joos]@cea.fr}
\and
\'Ecole Normale Sup\'erieure de Lyon, CRAL, UMR 5574 du CNRS, Universit\'e Lyon I, 46 All\'ee d'Italie, 69364 Lyon cedex 07, France\\
\email{benoit.commercon@ens-lyon.fr}}

%\date{Received <date> / Accepted <date>}

\abstract 
{The formation of protostellar discs is severely hampered by magnetic braking, as long as magnetic fields 
remain frozen in the gas. The latter condition depends on the levels of ionisation that characterise the innermost regions 
of a collapsing cloud.}
{The chemistry of dense cloud cores and, in particular, the ionisation fraction is largely controlled by cosmic rays. 
The aim of this paper is to evaluate whether the attenuation of the flux of cosmic rays expected in the regions 
around a forming protostar is sufficient to decouple the field from the gas, thereby influencing the formation of centrifugally 
supported disc.}
{We adopted the method developed in a former study to compute the attenuation
of the cosmic-ray flux as a function of the column density and the field strength in
clouds threaded by poloidal and toroidal magnetic fields.
We applied this formalism to models of low- and high-mass star formation extracted 
from numerical simulations of gravitational collapse that include rotation and turbulence.}
{For each model we determine the size of the magnetic decoupling zone, where collapse or rotation motion 
becomes unaffected by the local magnetic field. In general, we find that decoupling only occurs when
the attenuation of cosmic rays is taken into account with respect to a calculation in which 
the cosmic-ray ionisation rate is kept constant.
The extent of the decoupling zone also depends on the dust grain size distribution and is larger 
if large grains (of radius $\sim 10^{-5}$~cm) are formed by compression
and coagulation during cloud collapse. The decoupling region disappears for the high-mass
case. This is due to magnetic field diffusion caused by turbulence
that is not included in the low-mass models.}
{We conclude that a realistic treatment of cosmic-ray propagation and attenuation during cloud collapse
may lead to a value of the resistivity of the gas in the innermost few hundred AU around a forming protostar
that is higher than generally assumed.
Forthcoming self-consistent calculations should investigate whether this effect is strong enough to effectively decouple 
the gas from the field and to compute the amount of angular momentum lost by infalling fluid particles when they 
enter the decoupling zone.}


\maketitle

\section{Introduction}
\label{intro}

The study of the formation of circumstellar discs around protostars
still presents considerable theoretical challenges.  Because of the presence
of magnetic fields in the parent molecular clouds and cores
(Crutcher~\cite{c12}), assuming the strict conservation of angular momentum during cloud
collapse and star formation is not warranted. In fact, according to recent numerical and analytical 
studies, the main
effect of a magnetic field entrained by a collapsing cloud is to
brake any rotational motion, at least as long as the field remains frozen
in the gas and the rotation axis of the cloud is close to the mean
direction of the field (e.g. see Galli et al.~\cite{gl06}, Mellon \&
Li~\cite{ml08}, Hennebelle \& Fromang~\cite{hf08}).  However, discs
around Classes I and II young stellar objects are commonly observed
(e.g. Williams \& Cieza~\cite{wc11}, Takakuwa et al.~\cite{ts12}),
and there is also some evidence of discs around Class 0 objects (Tobin
et al.~\cite{th12}, Murillo et al.~\cite{mu13}).  

Different mechanisms
have been invoked as alleviating the problem of magnetic
braking during cloud collapse:
({\em i}\/) non-ideal magnetohydrodynamic (MHD) effects
(Shu et al.~\cite{sg06}, Dapp \& Basu~\cite{db10}, Krasnopolsky et
al.~\cite{kl11}, Braiding \& Wardle~\cite{bw12a,bw12b});
({\em ii}\/) misalignment between the main magnetic field direction and the rotation axis
(Hennebelle \& Ciardi~\cite{hc09}, Joos et al.~\cite{jh12}); 
({\em iii}\/) turbulent diffusion of the magnetic field (Seifried et al.~\cite{sb12},
Santos-Lima et al.~\cite{sd13}, Joos et al.~\cite{jh13});
({\em iv}\/) flux redistribution driven by the interchange instability (Krasnopolsky et al.~\cite{kl12});
and ({\em v}\/) depletion of the infalling envelope anchoring the magnetic field (Mellon \&
Li~\cite{ml09}, Machida et al.~\cite{mi11}).

Non-ideal MHD effects, namely ambipolar, Hall, and Ohmic diffusion,
depend on the abundances of charged species and on their mass and
charge. The ionisation fraction, in turn, is controlled by cosmic rays (hereafter CRs)
in cloud regions of relatively high column density 
(visual extinction $A_v\gtrsim 4$, McKee~\cite{m89}) 
where star formation takes place.
The CR ionisation rate is usually assumed to be equal to a
``standard'' (constant) value of $\zeta^{{\rm H}_2}\approx 10^{-17}$~s$^{-1}$,
often called the ``Spitzer'' value
(Spitzer \& Tomasko~\cite{st68}). However, CRs interacting with $\mathrm{H_{2}}$ 
in a molecular cloud lose energy by several processes, mainly
by ionisation losses (see Padovani et al.~\cite{pgg09}, hereafter PGG09).
As a consequence, while low-energy CRs ($E\lesssim100$~MeV)
are possibly prevented from entering a molecular cloud because of streaming instability
(Cesarsky \& V\"olk~\cite{cv78}), high-energy CRs are slowed down
to energies that are relevant for ionisation (ionisation cross sections for protons and electrons colliding with
$\mathrm{H_{2}}$ peak at about 100~keV and 0.1~keV, respectively).  

PGG09 show that
\zhh\ can decrease by about two orders of magnitude from 
``diffuse'' clouds of column density $\sim 10^{21}$~cm$^{-2}$ to 
``dense'' clouds and massive envelopes with column densities of $\sim 10^{24}$~cm$^{-2}$.
A similar attenuation can in principle take place during the process of 
cloud collapse, resulting in a decrease in \zhh\  in the inner region of a core where the
formation of a protostellar disc is expected to occur. A further attenuation is predicted 
by the toroidal field component generated by rotation
that increases the particles' path length and 
enhances the losses by magnetic mirroring (Padovani \& Galli~\cite{pg11}, hereafter PG11). 
A reduced CR ionisation rate results
in a more efficient ambipolar diffusion, which may help to alleviate the magnetic braking
problem (Mellon \& Li~\cite{ml09}). The aim
of this paper is to evaluate how variations in \zhh\ can affect 
the resistivity of the gas, and,
eventually, how CRs influence the dynamics of collapse and the formation of a circumstellar disc.

A full treatment of this problem, in which the propagation of CRs is computed self-consistently 
with the evolution of density and magnetic field, 
following at the 
same time the formation and destruction of chemical species, would be 
prohibitively time-consuming from a numerical point of view. Our approach is therefore
a simpler one. First, ({\em a}\/) we sacrifice the self-consistency by taking snapshots at particular 
evolutionary times of magnetic 
field configurations and density distributions from ideal MHD simulations that do not include any
resistivity of the gas. Second, ({\em b}\/) we propagate CRs in these configurations, and compute the spatial 
distribution of the CR ionisation rate. Then ({\em c}\/) we build a simplified chemical model (that can be 
used as a fast subroutine in any dynamical code) to approximately evaluate the chemical composition 
at each spatial position using as input the distribution of CR ionisation rates determined at the previous step.
Finally, ({\em d}) we compute the microscopic resistivities (ambipolar, Hall, and Ohmic) and compare the 
time scale of magnetic field diffusion $t_B$ to the dynamical time scale $t_{\rm dyn}$ 
at each point in the model to determine the region of magnetic decoupling, where $t_B <t_{\rm dyn}$.
In this region, the dynamics of the gas is unimpeded by magnetic forces, and whatever angular momentum
is carried by fluid particles crossing its border, it will be conserved in its interior. In particular, centrifugally 
supported discs can form inside this region (but not outside).
Clearly, the assumption of ideal MHD on which the simulations are based becomes invalid (by definition) 
inside the decoupling region. In this sense, our calculation is not fully self-consistent. However, our aims
in this preliminary investigation are: first, to show whether the extent of the decoupling region can be sufficiently 
large to allow the formation of a realistic disc (at least $\sim 10$~AU in radius); second, to compare the size 
of this region obtained with the accurate treatment of CR attenuation with column density and magnetic field
developed in our previous studies and with a constant value of \zhh, as usually assumed.

The paper is organised as follows. In Section~\ref{descmodel}, we
provide a description of the model of CR propagation as well as of
the numerical simulations adopted. In Section~\ref{ionfraction} we
describe the basic features of the chemical code that we employed
to compute the ionisation fractions, exploring the effect of different
grain size distributions on abundances. In Section~\ref{diffcoeff}
we calculate the diffusion coefficients, and, in Section~\ref{difftimes},
we use them to calculate the corresponding diffusion time scales.
Finally in Section~\ref{conclusions} we summarise our conclusions.
A full description of the chemical code is given in
Appendix~\ref{app:chemcode}.  The dependence of the diffusion
coefficients on the grain size distribution is examined in detail
in Appendix~\ref{app:meangrain}. Finally, in
Appendix~\ref{app:sepcontrdifftime} we describe the contribution
to the total diffusion time due to the different diffusion processes.

\section{Description of the models}
\label{descmodel}

PG11 showed that the magnetic fields of dense 
molecular cloud cores (modelled as equilibrium configurations) can influence the 
penetration and propagation of CRs from the intercloud medium to the core centre,
affecting the spatial distribution of the ionisation rate.
In fact, since charged particles spiral around
field lines, CRs propagating in complex magnetic configuration ``see'' a higher column density
than particles moving on straight trajectories, and therefore suffer larger energy losses. 
In addition, Padovani et al.~(\cite{phg13}, hereafter PHG13) 
found that the mirroring effect becomes stronger when the toroidal
field component is larger than about 40\% of the total field, in
the central $300-400$~AU where density is higher than $10^9$~cm$^{-3}$.
This makes the CR ionisation rate, \zhh\ (hereafter we refer specifically to the 
ionisation of H$_2$), to drop well below
$10^{-18}$~s$^{-1}$ down to about $10^{-20}$~s$^{-1}$, roughly equal 
to the ionisation level arising from the decay of long- and short-lived radionuclides 
within protoplanetary discs (Umebayashi \& Nakano~\cite{un81}, Cleeves et al.~\cite{cabv13}).
PHG13 performed a numerical study of the propagation of CRs
in collapsing clouds using several snapshots of numerical simulations, adopting the formalism
developed in PG11. They found that the value of \zhh\ can be orders of magnitude lower
than the standard ``Spitzer'' value in the inner region of a core where the formation
of  a protostellar disc is expected to take place. From their numerical results, 
they derived an approximate analytical
expression to compute the CR ionisation rate taking both 
column density and magnetic field effects into account. This formula provides a fast and efficient
way to calculate the distribution of \zhh\ that can be
employed to estimate the ionisation fractions of charged particles
in order to determine the diffusion coefficients.

In the following, we compute the distribution of \zhh\ for
three different configurations obtained from numerical
simulations performed with the AMR code RAMSES\footnote{RAMSES
simulations were analysed using PyMSES (Labadens et al.~\cite{lc11}).}
(Teyssier~\cite{t02}, Fromang et al.~\cite{fh06})
with the goal to compute the diffusion time scales in the collapse region. 
The first two cases are 
low-mass models from Joos et al.~(\cite{jh12}), while the third one
represents a high-mass model from Commer\c con et al. (in prep.). 
Table~\ref{tab:models} summarises the parameters.
We assumed the dependence of \zhh\ on column density given by model
$\cM$ in PHG13 (see their Fig.~1). This model is obtained by adopting the CR interstellar proton and 
electron spectra
presented in Webber~(\cite{w98}) and Strong et al.~(\cite{sm00}), respectively.

\begin{table}[!h]
\center
\caption{Parameters of the simulations described in the text: 
non-dimensional mass-to-flux ratio $\lambda$, initial angle between
the magnetic field direction and the rotation axis $\alpha_{\rm B,J}$, 
time after the
formation of the first Larson's core $t$ (core formed in the centre of
the pseudo-disc with $n\gtrsim10^{10}$~cm$^{-3}$ and $r\sim10-20$~AU),
initial mass $M_{\rm in}$,
mass of the protostellar core $M_{\bigstar}$ and of the disc $M_{\rm disc}$.}
\begin{tabular}{ccccccc}
\hline
Case & $\lambda$ & $\alpha_{\rm B,J}$ & $t$ & $M_{\rm in}$ & $M_{\bigstar}$ & $M_{\rm disc}$\\% & Disc ?\\
 &         & [rad]   & [kyr] & $[M_{\odot}]$ & $[M_{\odot}]$ & $[M_{\odot}]$\\% & (Y$^{a}$ / N$^{b}$ / K$^{c}$)\\
\hline\hline
$L_1$ & 5  &  0  & 0.824 & 1 & -- & --\\% & N\\
$L_2$ & 5  &  $\pi$/2 & 10.756 & 1 & 0.46 & 0.28\\% & K\\
$H$ & $\sim2$ & no initial rotation & 6.000 & 100 & 1.24$^{a}$ & 0.87$^{a}$\\ % & $X$\\
\hline
\end{tabular}\\[2pt]
\noindent $^a$ This value refers to the densest fragment formed.
\label{tab:models}
\end{table}

\subsection{Low-mass models}
\label{low}

Models $L_1$ and $L_2$ have been already used by PHG13 to calculate 
maps of the CR ionisation rate (see their Figs.~7 and 10) and they represent two extreme situations:
one snapshot taken at the beginning of the simulation and one at the end ($L_{1}$ and $L_{2}$, respectively).
These ideal MHD numerical simulations 
describe the collapse of a rotating core with initial mass of $1~M_{\odot}$ and a density profile given by
a modified power-law, 
\be
n(r)=\frac{n_0}{1+(r/r_0)^2}\,,
\ee\noindent
where $n_{0}=7.8\times 10^6$~cm$^{-3}$ 
and $r_0=4.68\times 10^{-3}$~pc according to observations
(Andr\'e et al.~\cite{awt00}, Belloche et al.~\cite{ba02}). The ratio of 
thermal-to-gravitational energy is about 0.25 and the
ratio of rotational-to-gravitational energy ratio is about 0.03. 
Model $L_1$ is a parallel rotator, i.e. the rotation axis is aligned to the magnetic 
axis, defined as the average direction of the field ($\alpha_{\rm B,J}=0$). 
No rotationally supported disc
has formed at the time of this particular snapshot 
($t=0.924$~kyr after the formation of the first Larson's core). In this case,
\zhh\ decreases to values of $2-4\times10^{-18}$~s$^{-1}$ in
the inner $100-200$~AU radius (see Fig.~7 in PHG13).  Model $L_2$
is a perpendicular rotator ($\alpha_{\rm B,J}=\pi/2$) and a late-time
configuration ($t=10.756$~kyr), showing a keplerian disc perpendicular 
to the rotation axis. Here \zhh\ drops to very low values, down to
$2\times10^{-21}$~s$^{-1}$ in the inner few tens of AU and it is
lower than $10^{-18}$~s$^{-1}$ in a region of about 200~AU radius
(see Fig.~10 in PHG13).  Clearly, at these low levels,
CRs compete with short-lived radionuclides in determining the ionisation fraction
(see e.g. Cleeves et al.~\cite{cabv13}).

\subsection{High-mass model}
\label{high}

Model $H$ has similar initial conditions as in Commer\c con et al.~(\cite{chh11}). It consists in a $100~M_{\odot}$ 
uniform temperature ($T=10$~K) dense core, with the same density profile shape as in models $L_{1}$ and 
$L_{2}$ ($n_{0}=1.2\times10^{7}$~cm$^{-3}$, $r_{0}=1.87\times10^{-2}$~pc, 
and a factor of 10 in density contrast between the centre and the border of the sphere). The turbulent-to-gravitational energy ratio is about 0.2, corresponding to an initial Mach number of 7, and the thermal-to-gravitational energy ratio is about 0.01. 
Following Hennebelle et al.~(\cite{hc11}) and Commer\c con et al.~(\cite{chh11}), 
we apply initial perturbations to the velocity field only to account for initial turbulence in the core, which is not 
driven at large scales after the start of the calculations. There is no global initial rotation in the model, 
namely the angular momentum is built from the initial velocity fluctuations. The initial magnetic field is aligned with the $x$-axis and its intensity is proportional to the column density through the cloud. 

\begin{figure}[!h] 
\begin{center}
\includegraphics[width=.5\textwidth]{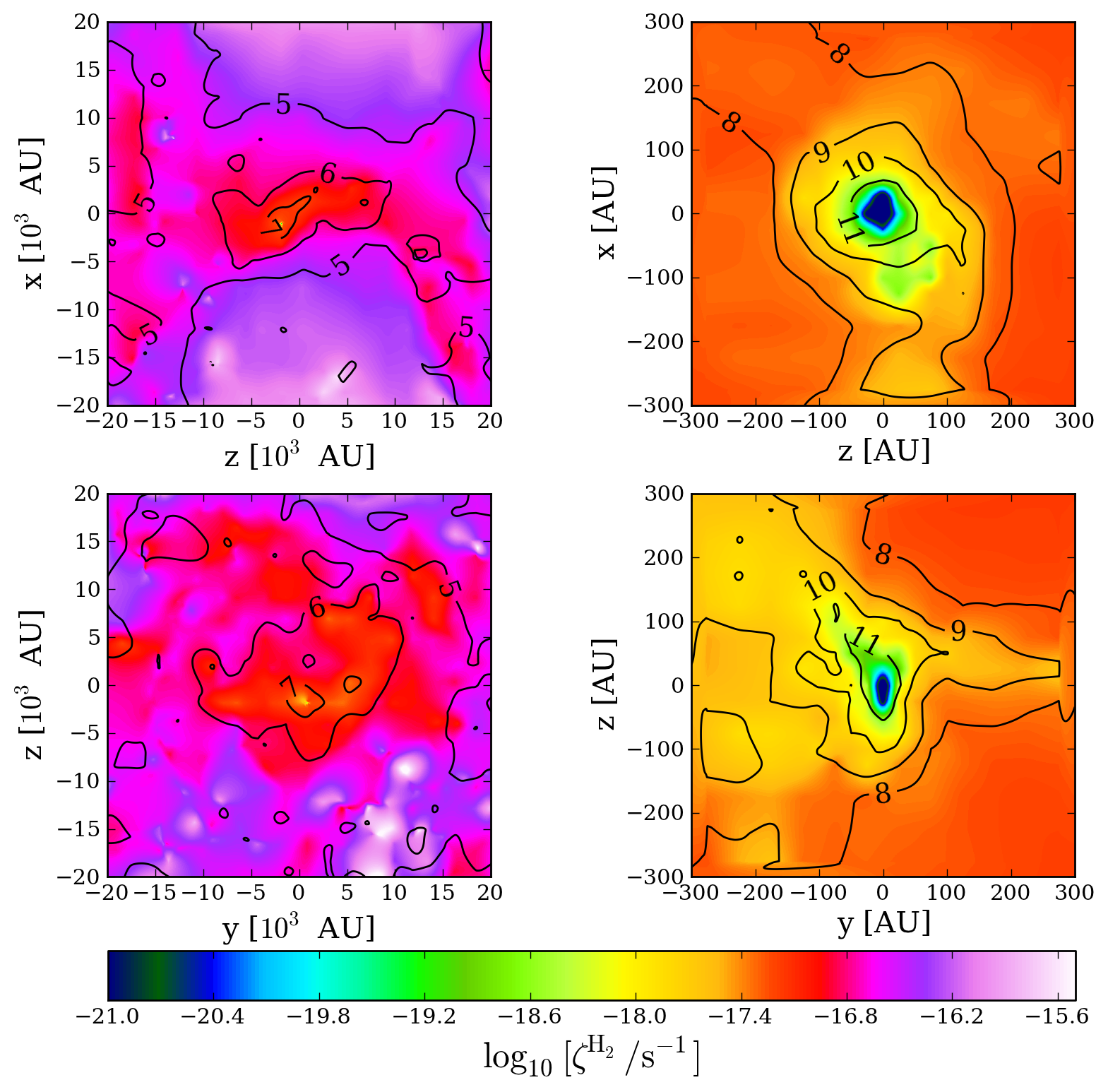}
\caption[]{CR ionisation maps and iso-density contours ({\em black
solid lines}) for the case $H$ in Table~\ref{tab:models}.  {\em Left}
panels show the entire computational domain while {\em right} panels
show a zoom in the inner region.  {\em Upper} and {\em lower} panels
show two perpendicular planes both containing the density peak.
Labels show $\log_{10}\ [n/\mathrm{cm^{-3}}]$.}
\label{RAMSES_MASS_M_0.7_0_145} 
\end{center} 
\end{figure}

In the following we focus on the densest fragment formed in the calculations whose properties are summarised in Table~\ref{tab:models} and we discuss the results only for the $(z,x)$ plane,
since it is close to the disc plane.  As model $L_2$, model $H$
shows $\Ezhh\simeq10^{-21}$~s$^{-1}$ in the inner few tens of AU,
but in the latter case the region with $\Ezhh\lesssim10^{-19}$~s$^{-1}$
is even larger, with a mean radius of 150~AU.  At a radius of 300~AU \zhh\ increases
only to up to $5\times10^{-18}$~s$^{-1}$ , while at
the same radius model $L_2$ has already reached CR ionisation
rate values larger than $10^{-17}$~s$^{-1}$.  This can be explained
by noting the larger extent of the region with high density in the high-mass
case with respect to the low-mass cases (see Fig.~\ref{RAMSES_MASS_M_0.7_0_145}).

\section{Abundances of charged species}
\label{ionfraction}

To compute the ionisation fraction we adopt a ``minimal'' chemical network 
(a simplified version of more extensive networks, like e.g. in Umebayashi \& Nakano~1990)
that computes the steady-state abundance of $\rm H^+$, $\rm H_3^+$, a typical molecular
ion $m$H$^+$ (e.g. HCO$^+$), a typical ``metal'' ion $M^+$ (e.g. Mg$^+$), 
electrons and dust grains as a function
of the H$_2$ density, temperature and CR ionisation rate \zhh\ at each spatial 
position in our models. For each species $i$, the abundance is given by $x(i)\equiv
n(i)/n({\rm H_2})$.  Neutral hydrogen is
assumed to be in the form of ${\rm H}_2$, namely $n({\rm H})=2n({\rm
H}_2)$. The fraction of charged vs. neutral grains 
is computed in a simplified way: if the density of
electrons is higher than the density of grains,
all grains are assumed to carry one electron; otherwise, the fraction
of negatively charged grains is determined by charge balance with
the positively charged species, and the residual number of grains
is assumed to be neutral. Positively charged grains as well as 
multiple charged grains are ignored for simplicity.
A detailed description of the chemical network adopted and a summary of the 
reactions included is given in
Appendix~\ref{app:chemcode}. 

As a benchmark test, we compared our results with those obtained 
with the publicly available code
ASTROCHEM\footnote{\tt http://smaret.github.io/astrochem/}, that includes about
a thousand of reactions,  
finding comparable results within a factor of 2 to 5. The use of our simplified
code is justified by the fact that it allows the calculation of the
ionisation fractions for each point of our models, 
in about 0.7 ms on average against about 20 s needed by ASTROCHEM.
Thus, our minimal chemical network, combined with the fitting
formula given by Eqs. (19)--(24) of PHG13 to compute \zhh$(N,{\bf
B})$, is an effective tool to rapidly compute the fractional
abundances and diffusion coefficients at each time step
in non-ideal MHD simulations.

\subsection{Effects of the grain size distribution}
\label{grainsize}

Grains play a decisive role in determining the degree of coupling between
the gas and the magnetic field. In fact, the electrical resistivity of the gas depends on the
abundance and the size distribution of charged grains: larger grains have
a smaller Hall parameter (see Sect.~\ref{diffcoeff}) than smaller grains, and therefore are 
less coupled to the magnetic field.
To cover all possible situations,
we run our chemical model for three different grain size distribution.
In particular, we fix the maximum size, $a_{\rm
max}=3\times10^{-5}$~cm (Nakano et al.~\cite{nn02}), while we vary
the minimum grain size: $(i\/)$ $a_{\rm min}=10^{-5}$~cm, representative of
large grains formed by compression and coagulation during the collapse 
(Flower et al.~\cite{fp05}); $(ii\/)$ $a_{\rm
min}=10^{-6}$~cm, the minimum grain radius of a MRN size
distribution (Mathis et al.~\cite{mrn77}) that gives the same grain opacity found by
Flower et al.~(\cite{fp05}); and $(iii\/)$ $a_{\rm min}=10^{-7}$~cm,
a typical size for very small grains.

\begin{figure}[]
\begin{center}
\includegraphics[width=.45\textwidth]{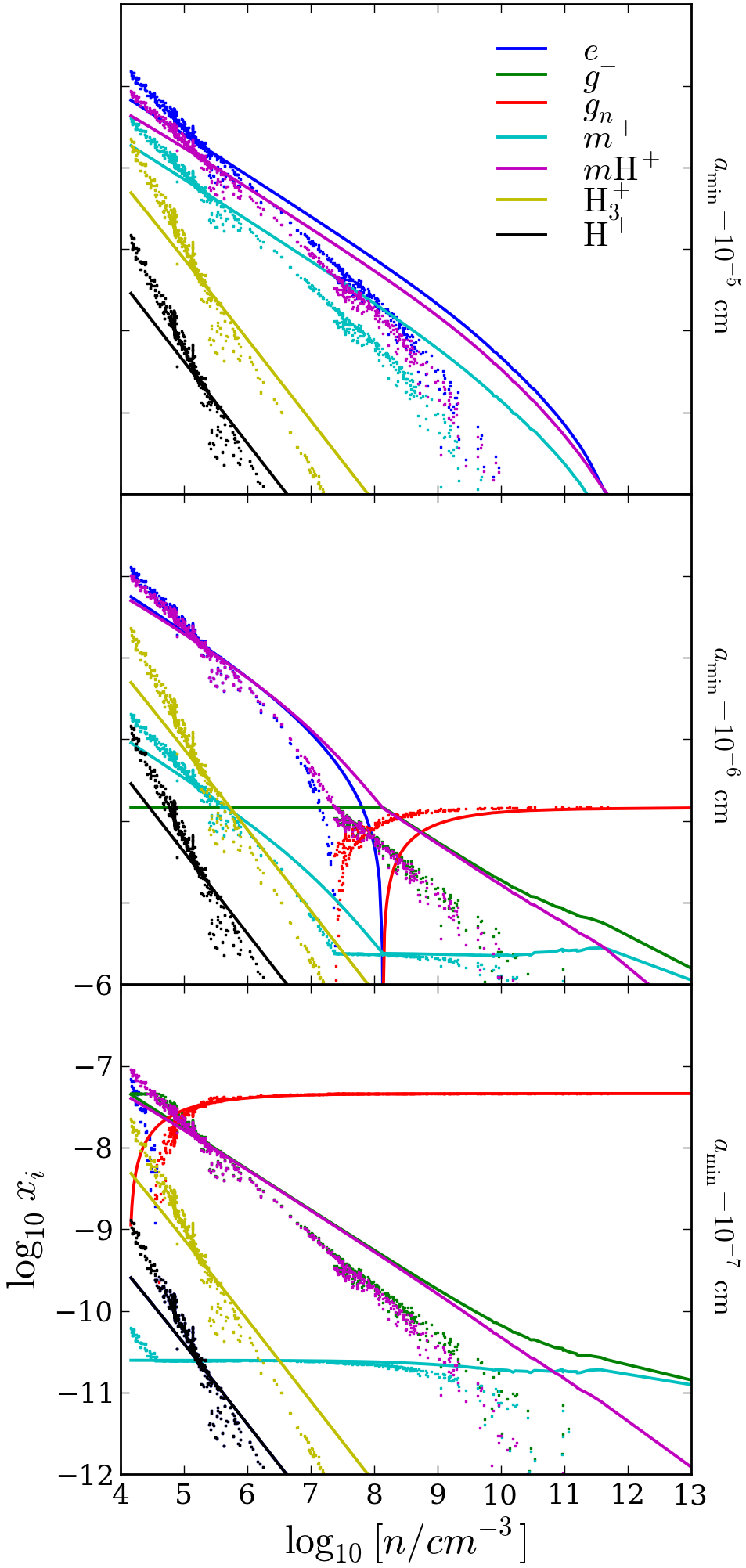}
\caption[]{Ionisation fractions for model $L_2$ as a function of the volume density
computed with a constant CR ionisation rate \zhh$=5\times10^{-17}$~s$^{-1}$ 
(constant-$\zeta$, {\em solid lines})
and with a spatially resolved \zhh\ (variable-$\zeta$, {\em dots}). 
The three panels are for $a_{\rm min}=10^{-5}$~cm ({\em top}\/),
$a_{\rm min}=10^{-6}$~cm ({\em middle}\/) and $a_{\rm min}=10^{-7}$~cm ({\em bottom}\/).}
\label{ionfrac}
\end{center}
\end{figure}

In Fig.~\ref{ionfrac}
we compare the abundances computed with our minimal chemical 
network assuming a constant
$\zeta^\mathrm{H_2}=5\times10^{-17}$~s$^{-1}$ (hereafter ``constant-$\zeta$'') 
with those obtained from the spatially-resolved 
values of \zhh\ computed by PHG13 (hereafter ``variable-$\zeta$''). The comparison is shown only for the model $L_2$ 
(models $L_1$ and $H$ give similar results).  As a general remark, independently of
$a_{\rm min}$, the variable-$\zeta$ model gives larger abundances of charged 
species than the constant-$\zeta$ model at $n\lesssim10^{6}$~cm$^{-3}$.
Conversely, at higher
densities, abundances from the variable-$\zeta$ model are well below 
those resulting from the constant-$\zeta$ model.
This happens because the variable-$\zeta$ model is higher than the constant-$\zeta$ at
$n\lesssim10^{6}$~cm$^{-3}$, while it quickly decreases below $\zeta^\mathrm{H_2}=5\times10^{-17}$~s$^{-1}$
at higher densities.

\subsection{Dependence of chemical abundances on \zhh}
\label{grainsize}

\begin{figure}[]
\begin{center}
\includegraphics[scale=.25]{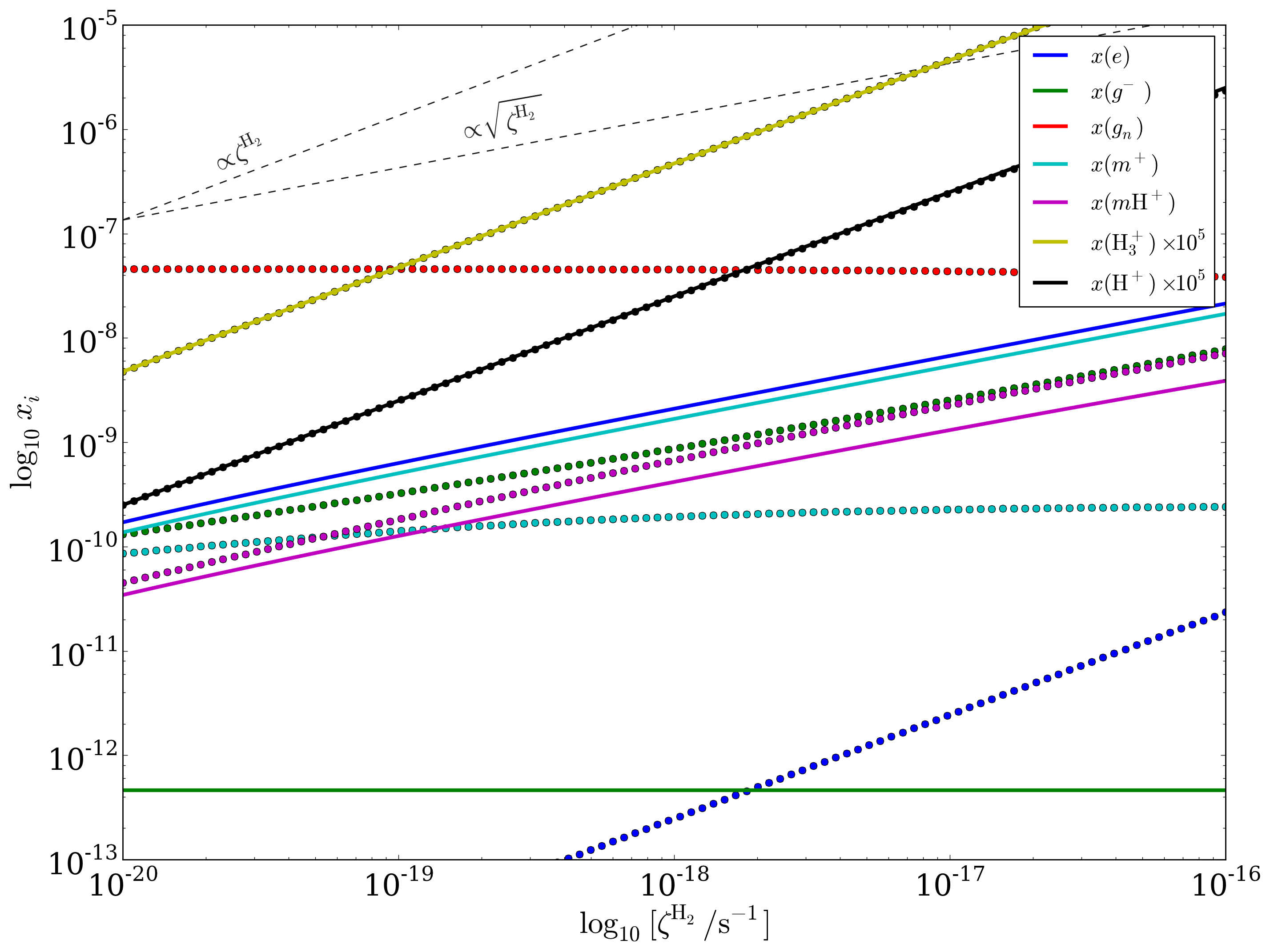}
\caption[]{Ionisation fractions as a function of the CR ionisation rate 
at density $n(\mathrm{H_2})=10^{6}$~cm$^{-3}$
assuming large grains ($a_{\rm min}=10^{-5}$~cm, {\em solid lines}) and
small grains ($a_{\rm min}=10^{-7}$~cm, {\em dotted lines}). {\em Dashed grey lines} show the
trend for $x_{i}\propto\Ezhh$ and $x_i\propto(\Ezhh)^{1/2}$.}
\label{ionfraction_vs_zeta}
\end{center}
\end{figure}

The relation between densities of charged species and neutrals is usually 
expressed in the form 
\be
\label{ninn}
n(i)\propto (\Ezhh)^{k^\prime}\,,
\ee
with $k^\prime\approx 1/2$
(e.g. see Ciolek \& Mouschovias~\cite{cm94,cm95}).
However, $k^\prime$ can differ from $1/2$ depending on the grain size 
(except for molecular ions). Fig.~\ref{ionfraction_vs_zeta} shows the dependence of 
the chemical species computed with our minimal model as a function of \zhh.
As shown by the Figure, in the case of large grains the abundance of electrons and metal ions follows 
Eq.~(\ref{ninn}) with $k^\prime=1/2$, while negatively
charged grains are independent of \zhh. For small grain size
$k^\prime$ increases towards 1 and $1/2$ for electrons and negative
grains, respectively, whereas metal ions become independent of \zhh.
It is worth noting that in the case of strong depletion,
$\mathrm{H_3^+}$ and $\mathrm{H^+}$ are the most abundant
species and both $x(\mathrm{H_3^+})$ and $x(\mathrm{H^+})$
are proportional to \zhh\ irrespective of grain size and density.

\section{Diffusion coefficients}
\label{diffcoeff}

The electrical resistivity of a plasma is a measure of the ability of the 
magnetic field and the charges attached to it to move (diffuse) with respect to 
the neutrals and/or other charges. 
In the weakly ionised gas that characterise dense cores, the resistivity is dominated 
by ambipolar diffusion, a process by which charged particles develop a drift velocity
with respect to the neutral component, and the Lorentz force acting
on the charges is conveyed to the neutral gas through collisions.

For each charged species $i$ in a sea of neutrals (molecular hydrogen), the parameter gauging the relative importance
of the Lorentz and drag forces is the Hall parameter (e.g. Wardle
\& Ng~\cite{wn99}) defined by
\be
\label{Hpar}
\beta_{i,\mathrm{H_{2}}}=\left(\frac{Z_{i}eB}{m_{i}c}\right) 
\frac{m_{i}+m_\mathrm{H_{2}}}{\mu m_{\mathrm{H}}n(\mathrm{H_2})\langle\sigma v\rangle_{i,\mathrm{H_{2}}}}\,,
\ee 
where $m_{i}$ and $Z_{i}e$ are the mass and the charge of the species
$i$, respectively, and $\mu=2.36$ is the molecular weight for the
assumed fractional abundances of $\mathrm{H_{2}}$ and He.  The
momentum transfer rate coefficients $\langle\sigma
v\rangle_{i,\mathrm{H_{2}}}$ have been parameterised as a function of
temperature and relative speed in Pinto \& Galli~(\cite{pg08}).

Drifts of charged species with respect to neutrals determines
different regimes for the magnetic diffusivity. The induction equation then becomes
\begin{eqnarray}
\label{dBdt} 
\frac{\partial\vec B}{\partial
t}+\nabla\times(\vec B\times\vec U) 
&=& \nabla\times\left\{\eta_\mathrm{O}\nabla\times\vec
B+\eta_\mathrm{H}(\nabla\times\vec B)\times\frac{\vec B}{B}\right.\\\nonumber
& & +\left.\eta_\mathrm{AD}\left[(\nabla\times\vec B)\times\frac{\vec B}{B}\right]\times
\frac{\vec B}{B}\right\}\,, 
\end{eqnarray}
where $\vec U$ is the fluid velocity and $\vec B$ the magnetic field
vector. Ambipolar, Hall, and Ohmic resistivities (\ea, \eh, and
\eo, respectively) can be written as a function of the parallel
($\sigma_{\parallel}$), Pedersen ($\sigma_{\rm P}$) and Hall
($\sigma_{\rm H}$) conductivities (e.g. see Wardle~\cite{w07}, Pinto
et al.~\cite{pgb08})
\begin{eqnarray}
\eta_{\rm AD} &=& \frac{c^{2}}{4\pi}\left(\frac{\sigma_{\rm
P}}{\sigma_{\rm P}^{2}+\sigma_{\rm
H}^{2}}-\frac{1}{\sigma_{\parallel}}\right),
\label{eqetaAD} \\ 
\eta_{\rm H} &=& \frac{c^{2}}{4\pi}\left(\frac{\sigma_{\rm H}}{\sigma_{\rm
P}^{2}+\sigma_{\rm H}^{2}}\right),
\label{eqetaH} \\ 
\eta_{\rm O} &=& \frac{c^{2}}{4\pi\sigma_{\parallel}}\,,
\label{eqetaO}
\end{eqnarray}
which are defined by 
\begin{eqnarray} 
\sigma_{\parallel} &=&
\frac{ecn(\mathrm{H_2})}{B}\sum_{i}Z_{i}x_{i}\ \beta_{i,\mathrm{H_{2}}}\,,\\
\sigma_{\rm P} &=&
\frac{ecn(\mathrm{H_2})}{B}\sum_{i}\frac{Z_ix_i\beta_{i,\mathrm{H_{2}}}}{1+\beta_{i,\mathrm{H_{2}}}^{2}}\,,\\
\sigma_{\rm H} &=&
\frac{ecn(\mathrm{H_2})}{B}\sum_{i}\frac{Z_i x_i}{1+\beta_{i,\mathrm{H_{2}}}^{2}}\,.
\end{eqnarray}

\begin{figure}[] \begin{center}
\includegraphics[width=0.5\textwidth]{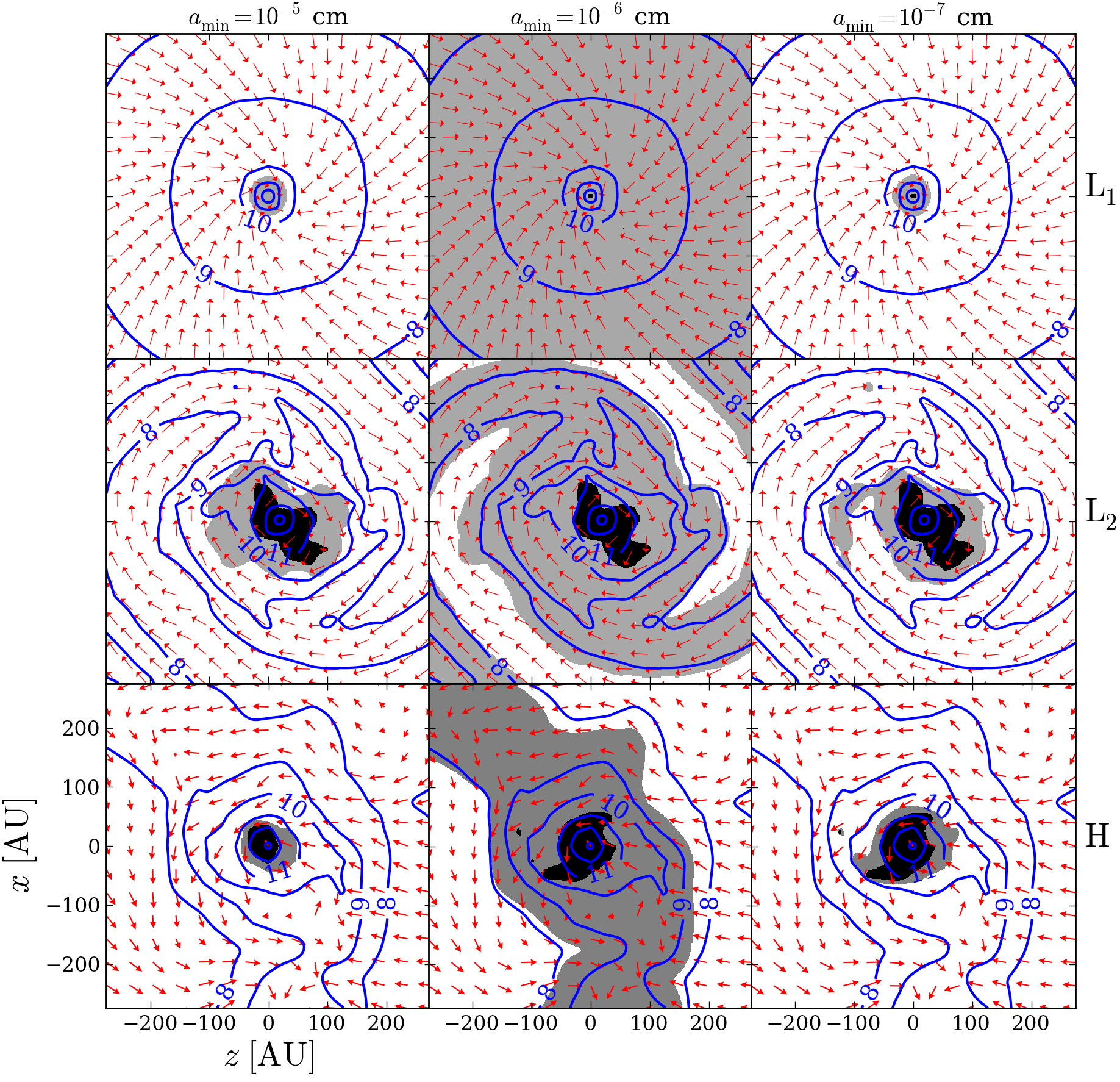}
\caption[]{Density contours ({\em solid blue lines} with labels
indicating $\log_{10}[n/\mathrm{cm^{-3}}]$) superposed to the velocity
field ({\em red arrows}). The shaded areas show regions dominated
by ambipolar ({\em white}), Hall ({\em grey}), and Ohmic ({\em
black}) diffusion. The resistivities are calculated for three different 
values of $a_{\rm min}$ and for the three models $L_1$, $L_{2}$, and $H$.} 
\label{dominant_diffusion_3x3_paper}
\end{center} \end{figure}

In general, the ambipolar resistivity term controls diffusion in
low density regions ($n_\mathrm{H_{2}}\lesssim10^{8}-10^{9}$~cm$^{-3}$),
whereas Hall diffusion dominates at intermediate densities
($10^{8}-10^{9}$~cm$^{-3}\lesssim
n_\mathrm{H_{2}}\lesssim10^{11}$~cm$^{-3}$) 
and Ohmic dissipation sets in 
at even higher densities ($n_\mathrm{H_{2}}\gtrsim10^{11}$~cm$^{-3}$), see Umebayashi \& Nakano~(\cite{un81}).
However, the extent to which a diffusion process dominates over the
others hinges on several factors, one of which is the assumed grain size distribution. 
To show this effect, we computed the ionisation fractions using the values
of \zhh\ for the three models ($L_1$, $L_2$, and $H$) and we
compared the spatial distribution of the corresponding resistivities
varying $a_{\rm min}$.  Background colours in
Fig.~\ref{dominant_diffusion_3x3_paper} show the predominant diffusion
mechanism.  The distributions with
$a_{\rm min}=10^{-5}$ and $10^{-7}$~cm lead to similar results, independently on the
distribution of \zhh. This can be explained by looking at the trend
of the different resistivity contributions as a function of the grain size
and $n({\rm H}_2)$ (see Appendix~\ref{app:meangrain}). In
fact, the ambipolar diffusion term does not increase monotonically with 
grain radius but, at densities of $10^{8}-10^{9}$~cm$^{-3}$, it
shows an absolute minimum around $a_{\rm min}=10^{-6}$~cm, while
the Hall term continues to grow. At the highest densities 
the Ohmic term prevails, except for model $L_1$.
The Hall term
can play an important role down to densities of about $10^{8}$~cm$^{-3}$
for $a_{\rm min}=10^{-6}$~cm.

\section{Diffusion time scales}
\label{difftimes}

The drift velocity of the magnetic field can be represented by the
velocity of the charged species, which are frozen with field lines,
with respect to neutrals. From the comparison of this velocity with
the fluid velocity, it is possible to assess the degree of diffusion
of the field and then to estimate the size of the region where gas
and magnetic field are decoupled. In principle, a more direct estimate of the reduction
of magnetic braking resulting from a decrease in the CR ionisation rate
would come from a comparison between the magnetic braking time with 
the dynamical time of the flow. However, a simple estimate of the former is not 
straightforward, as it depends crucially on the field morphology, strength, and 
relative orientation of the average field direction and the cloud's angular momentum.
For this reason, we prefer to present a comparison, at each spatial position
at a given time step, between the diffusion time of the field and the dynamical time
of the flow as defined in the following.

The magnetic field drift velocity $\vec U_{B}$ can be written as
a function of resistivities (Nakano et al.~\cite{nn02}), 
allowing to isolate the ambipolar (AD), Hall (H), and Ohmic (O)
contributions, namely
\be 
\vec U_{B}=\vec U_{\rm AD}+\vec U_{\rm H}+\vec U_{\rm O}\,, 
\ee
where 
\begin{eqnarray} 
\vec U_{\rm AD} &=& \frac{4\pi\ \eta_{\rm
AD}}{cB^{2}}\vec j\times\vec B\,,\\ \vec U_{\rm H} &=& \frac{4\pi\
\eta_{\rm H}}{cB^{3}}(\vec j\times\vec B)\times\vec B\,,\\ \vec
U_{\rm O} &=& \frac{4\pi\ \eta_{\rm O}}{cB^{2}}\vec j\times\vec B\,
\end{eqnarray} 
and 
\be 
\vec U_B=\frac{4\pi}{cB^{2}}\left[\left(\eta_{\rm AD}+\eta_{\rm
O}\right)\vec j\times\vec B+\eta_{\rm H}\left(\vec j\times\vec
B\right)\times\frac{\vec B}{B}\right]\,,  
\ee 
where $\vec j=(c/4\pi)\nabla\times\vec B$ is the current density.
Thus, the diffusion time of the magnetic field,
$t_B$, can be written as a function of the time scales associated to the three diffusion
processes, 
\be 
\frac{1}{t_B}=\frac{1}{t_{\rm AD}}+\frac{1}{t_{\rm H}}+\frac{1}{t_{\rm O}}\,,
\ee 
where $t_k=R/U_k$ ($k=\mathrm{AD,H,O}$) and $R$ a typical length scale of the region
(in the following at each point we take $R$ equal to the distance from the density peak).  
The diffusion time of the magnetic field can 
then be compared to the time scale of evolution of the fluid (for example, Nakano et al.~\cite{nn02}
compare $t_B$ to the free-fall time of a spherical homogeneous cloud). 
In this work we define the dynamical time scale of the cloud as $t_{\rm dyn}=R/U$, where $U$
is the fluid velocity, including both infall and rotation. 
In regions where $t_{B}<t_{\rm dyn}$  
the magnetic field is partially decoupled and therefore has less influence on the gas dynamics while, if $t_{B}>t_{\rm
dyn}$, diffusion is not efficient enough and the magnetic field remains well
coupled to the gas.
To stress the importance of properly taking the propagation 
of CRs inside a molecular cloud into account, we also compare the dynamical time with the diffusion time
computed in the constant-$\zeta$ case assuming \zhh$=5\times10^{-17}$~s$^{-1}$ 
(hereafter $t_{B,\zeta-{\rm const}}$), and in the variable-$\zeta$ case, taking \zhh\ from PHG13 (hereafter
$t_{B,\zeta-{\rm var}})$.

\subsection{Low-mass models}

\begin{figure}[]
\begin{center}
\includegraphics[width=0.45\textwidth]{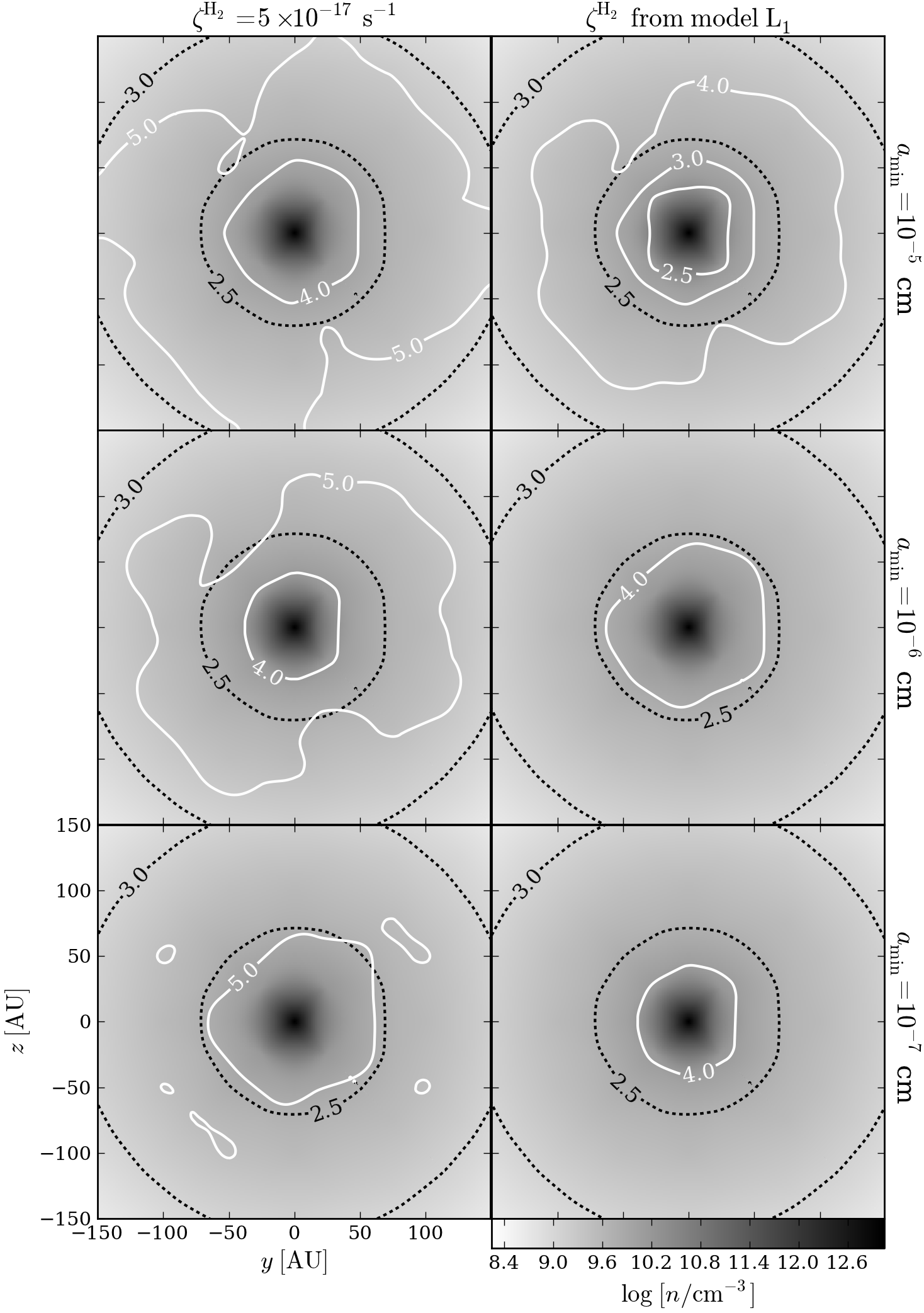}
\caption[]{Diffusion time contours ({\em white solid lines}) evaluated
with constant \zhh$=5\times10^{-17}$~s$^{-1}$ ({\em left column})
and \zhh\ from model $L_1$ in Table~\ref{tab:models} ({\em right
column}) compared with dynamical time contours ({\em black dashed
lines}) for three different values of the minimum grain size. Labels
show $\log_{10}(t/{\rm yr})$.}
\label{drifttime_plotpaper_YZ_100}
\end{center}
\end{figure}

Figure~\ref{drifttime_plotpaper_YZ_100} shows contours of
diffusion and dynamical time scales for the aligned rotator model $L_1$. Although the disc
is not formed at this early stage ($t=0.824$~kyr), %in this case, 
for $a_{\rm min}=10^{-5}$~cm in the variable-$\zeta$ case the magnetic diffusion time 
becomes shorter than the dynamical time in a central region with a radius
of about 20~AU. For smaller grains,
$t_{B,\zeta-{\rm var}}$ is always larger than $t_{\rm dyn}$ and no decoupling zone is 
formed. The diffusion time computed in the constant-$\zeta$ case is about one 
order of magnitude longer.
Even at later times ($t=11.025$~kyr), Joos et al.~(\cite{jh12}) find no disc formation for this 
aligned rotator simulation. However,  the low ionisation rate found by PHG13 in a small 
central region of density higher than 
$\sim 10^9$~cm$^{-3}$ (see their Fig.~8) can presumably promote the formation of a centrifugally supported 
disc via enhanced Ohmic and Hall diffusion, at least in the case of large grains.

\begin{figure}[]
\begin{center}
\includegraphics[width=0.45\textwidth]{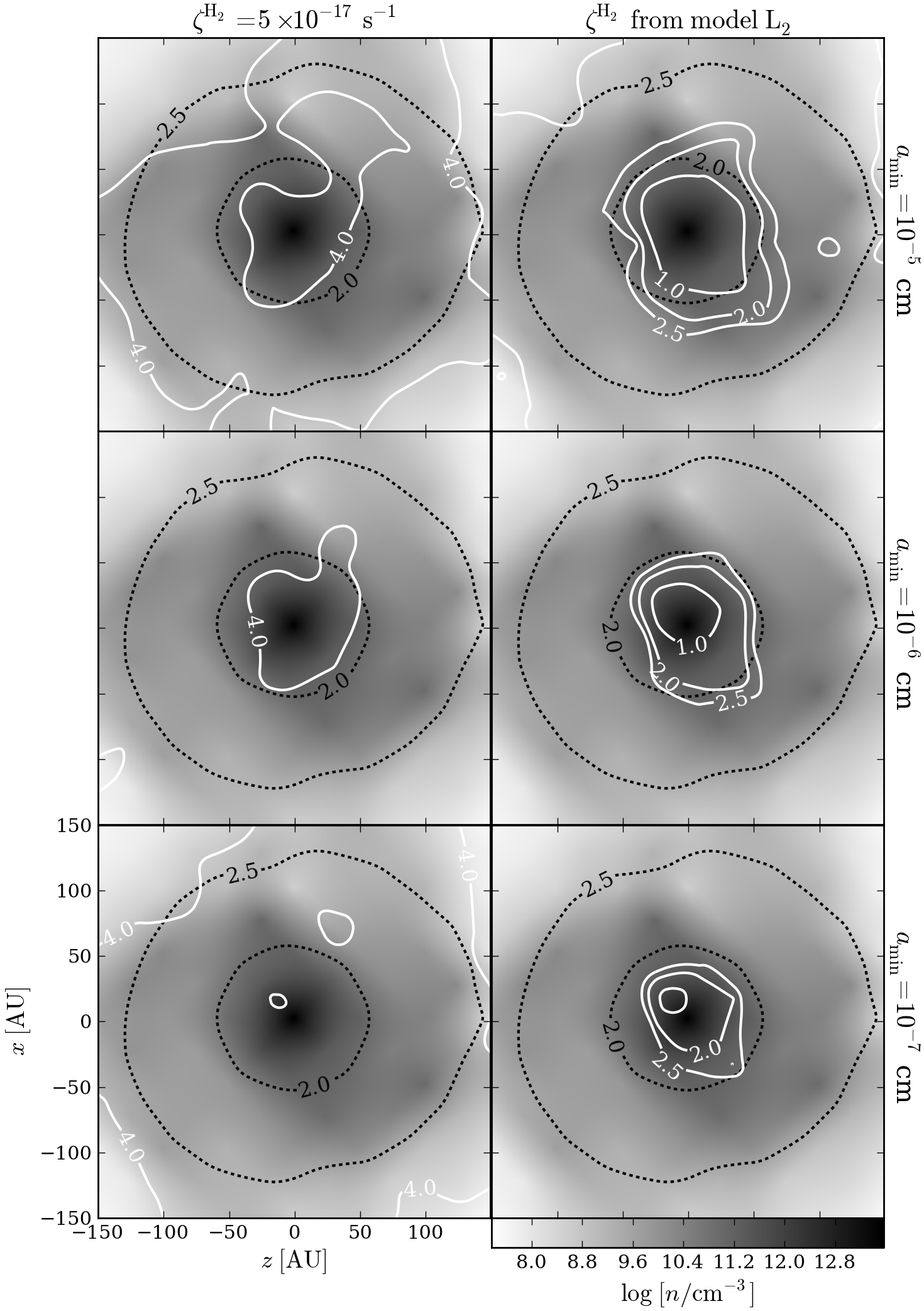}
\caption[]{Diffusion time contours ({\em white solid lines}) evaluated
with constant \zhh$=5\times10^{-17}$~s$^{-1}$ ({\em left column})
and \zhh\ from model L$_{2}$ in Table~\ref{tab:models} ({\em right
column}) compared with dynamical time contours ({\em black dashed
lines}) for three different values of the minimum grain size. Labels
show $\log_{10}(t/{\rm yr})$.}
\label{drifttime_plotpaper_ZX_187}
\end{center}
\end{figure}

A more interesting case is shown by the model $L_2$
(Fig.~\ref{drifttime_plotpaper_ZX_187}), where a protostellar disc of radius $\sim 200$~AU
has formed. While $t_{B,\zeta-{\rm
const}}$ is always larger than $t_{\rm dyn}$, $t_{B,\zeta-{\rm var}}$
is lower than $t_{\rm dyn}$ inside a decoupling zone whose size depends
on $a_{\rm min}$.  Magnetic decoupling is favoured by
large grains ($a_{\rm min}=10^{-5}$~cm) that are expected to form
by coagulation of smaller grains during the collapse.  In particular,
the lower left panel shows that in the inner region with a radius
of about 50~AU the gas experiences collapse while the magnetic
field diffuses. The region of decoupling shrinks with decreasing
$a_{\rm min}$, but even with the smallest $a_{\rm min}=10^{-7}$~cm
the decoupling occurs inside a region of about 20~AU of radius.
This comparison definitely proves that CRs play a crucial role in
determining the protostellar collapse time scale. In fact, the correct
evaluation of the CR ionisation rate as a function of density and
magnetic field allows the diffusion time to decrease up to three order
of magnitudes.  
In Appendix~\ref{app:sepcontrdifftime} we also show
separately the contribution to $t_B$ resulting from ambipolar, Hall, and
Ohmic diffusion, respectively.

\subsection{High-mass model}

\begin{figure}[]
\begin{center}
\includegraphics[width=0.45\textwidth]{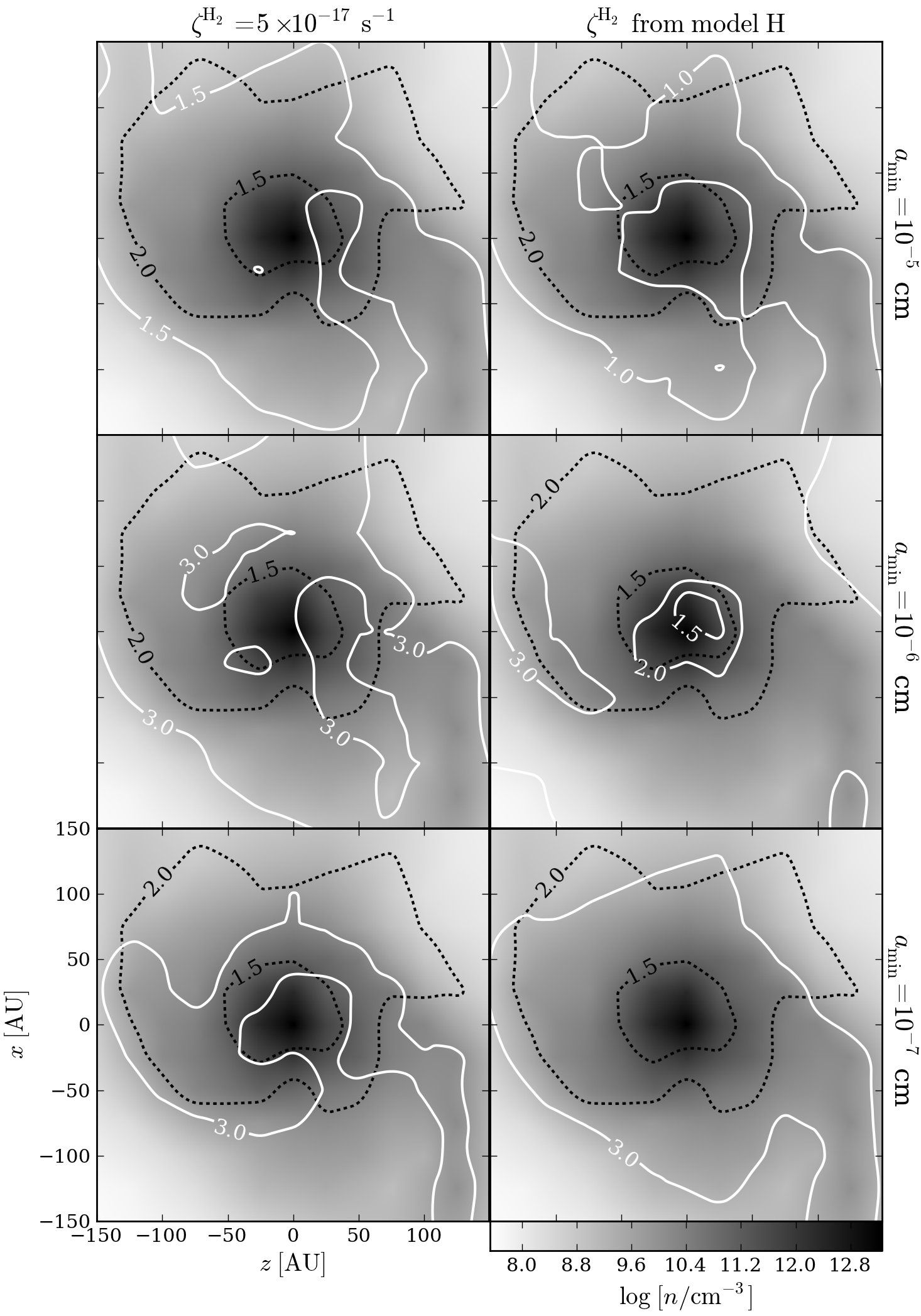}
\caption[]{Diffusion time contours ({\em white solid lines}) evaluated
with constant \zhh$=5\times10^{-17}$~s$^{-1}$ ({\em left column})
and \zhh\ from model H in Table~\ref{tab:models} ({\em right column})
compared with dynamical time contours ({\em black dashed lines}) for
three different values of the minimum grain size. Labels show
$\log_{10}(t/{\rm yr})$.}
\label{drifttime_plotpaper_ZX_145}
\end{center}
\end{figure}

The high-mass case (Fig.~\ref{drifttime_plotpaper_ZX_145}) shows a
similar behaviour to model $L_2$, but in the case of $a_{\rm
min}=10^{-5}$~cm, the decoupling between gas and field is allowed
in a even larger region, of about 100~AU of radius where both
$t_{B,\zeta-{\rm var}}$ and $t_{B,\zeta-{\rm const}}$ are lower than $t_{\rm
dyn}$. Decreasing the grain size, the region of decoupling becomes
narrower and vanishes for $a_{\rm min}=10^{-7}$~cm. The lack of a decoupling region 
in this high-mass case is at variance with the low-mass case discussed in the previous Section.
This depends on the field diffusing faster in the high-mass case owing to the turbulent nature of the flow,
as already stressed by Hennebelle et al.~(\cite{hc11}), Joos et al.~(\cite{jh13}). This diffusion is largely due to numerical
effects associated to the finite spatial resolution of the simulation. This ``numerical'' field diffusion results in a 
reduced strength of the magnetic field with respect to the (non-turbulent) low-mass case, and this in turn
makes the microscopic resistivity smaller (especially ambipolar diffusion, 
which is proportional to $B^2$). In this case, the microscopic resistivity is not sufficient to produce a 
significant decoupling region. Of course, if turbulence actually affects  
the field diffusion, the microscopic resistivity is not relevant.

%For the case of small grains, the different
%behaviour with respect to the low-mass
%case depends on the fact that even if the
%initial mass-to-flux ratio is close to 2, it increases at high
%density due to turbulent diffusion, although numerical effects certainly 
%affect this process (Hennebelle et al.~\cite{hc11}). 
%This enhanced diffusion produced by the turbulence
%being more effective than ambipolar diffusion, explains why in the
%case of small grains ($a_{\rm min}=10^{-7}$~cm) the diffusion time
%is larger than the dynamical time so that there is no decoupling as
%for the low-mass case.

\section{Conclusions}
\label{conclusions}

In this paper we investigated the role of CRs in the
formation of protostellar discs, using the results of PHG13 
to evaluate CR ionisation rate in collapsing low-mass clouds
accounting for both column density and magnetic effects. 
We developed a simple chemical code based on a minimal set of reactions 
to calculate, at each position in space at any given time, the abundances of the charged species, 
the corresponding electric resistivities, and 
the times scales associated to the main magnetic diffusion processes 
(ambipolar, Hall, and Ohmic). We applied this formalism to selected snapshots of 
numerical MHD simulations of the collapse of low- and high-mass clouds. 
Comparing the magnetic diffusion and the 
dynamical time scale, we determined the extent of the area where 
the gas is dynamically decoupled from the field. Inside this region, 
field freezing is invalid, and magnetic braking is ineffective. 

We performed our calculations in two cases: assuming a 
spatially uniform CR ionisation rate of $\Ezhh=5\times10^{-17}$~s$^{-1}$,
and adopting the formalism of PHG13 to evaluate the attenuation of 
CRs in a magnetised cloud. We found that the ionisation fraction is 
significantly lower in the second case, and therefore the coupling with 
the magnetic field is weaker than usually assumed in the central region
of a collapsing cloud. In particular:

\begin{enumerate}

\item For model $L_2$, a late-time snapshot of 
the collapse of a low-mass cloud with mean magnetic field perpendicular to the rotation 
axis, a decoupling zone of radius of $50-100$~AU around the central protostar is found in 
the case of variable \zhh, but not when \zhh\ is assumed constant. This size compares well with the 
size of protostellar discs. This stresses the importance of accounting 
for CR attenuation when computing ionisation fractions.

\item No decoupling zone is found in model $L_1$, an early time snapshot
of the collapse of a low-mass cloud with aligned field and rotation axis. In this model,
the time scale of field diffusion remains longer than the dynamical
time scale everywhere, either for a uniform or attenuated CR ionisation rate. In this case,
the relatively modest increase in the density and field strength by compression 
are insufficient to attenuate the CR flux to levels low enough to produce significant
decoupling.

\item A decoupling zone of size $\sim 100$~AU is also found in the case of model $H$, a snapshot from a 
typical high-mass collapse simulation. However, its size becomes smaller for smaller grains, and 
disappears altogether for grains smaller than $10^{-6}$~cm. 

\end{enumerate}

Although the models adopted do not represent a time sequence, they nevertheless 
suggest that a decrease in the ionisation and/or an increase in the resistivity occurs 
in the innermost region of a cloud some time 
after the onset of collapse, but not earlier. In fact, the conditions for a substantial increase in the magnetic 
diffusion time  are that the field is considerably twisted 
(compare the late-time misaligned model $L_2$ to the early-time 
aligned model $L_1$) and that dust grains had time to grow by coagulation (compare the upper and lower panels 
of Figs. 5--7). It is tempting to speculate that large, 100~AU-size discs are only allowed to form at a later stage when
the powerful magnetic brake on the infalling gas has been relieved by either (or a combination) of these effects.

Other results of our study are:

\begin{enumerate}

\item The dominant diffusion processes are ambipolar and Hall diffusion, with Ohmic 
resistivity becoming important only at the highest densities reached in our models
($n\gtrsim10^{11}$~cm$^{-3}$), where $\eta_{\rm O}$ is a factor of few larger than $\eta_{\rm AD}$ and $\eta_{\rm H}$,
(see black regions in Fig.~\ref{dominant_diffusion_3x3_paper}). 
In general, ambipolar and Hall diffusion are comparable 
over the innermost few hundred AU around a forming protostar. Thus, Hall diffusion
should not be neglected in non-ideal MHD collapse calculations. However, the region 
where a diffusion process dominates
over the others depends sensitively on the
grain size distribution.  In fact, while Hall and Ohmic resistivities
have a monotonic dependence on the grain size (all else being equal), 
ambipolar diffusion has
a minimum for a mean grain radius of about $1-3\times10^{-6}$~cm
especially at high densities.

\item In general, the size of the decoupling zone 
decreases for smaller grain size. The maximum grain size assumed here,
($a_{\rm min}=10^{-5}$~cm), is likely a realistic value for the condition
expected in disc-forming regions, where larger grains are predicted
to form by compression and coagulation of smaller grains. Both
Models $L_2$ and $H$, referring to low- and high-mass cases,
respectively, predict the formation of a protostellar disc, but for
model $H$ the decoupling region disappears for smaller grains ($a_{\rm
min}=10^{-7}$~cm), since the magnetic field becomes weaker. 

\item The dependence of the fractional abundances of charged species on
the CR ionisation rate may differ from the usually assumed $\sqrt{\Ezhh}$ dependence.
Our minimal chemical model shows that depending on the grain size distribution, 
the abundance of several species can scale as \zhh\ or become independent on \zhh,
and in general the dependence is not the same for all charged species.
Thus, care should be taken when the ionisation fraction is scaled with the CR flux.

\end{enumerate}

To summarise, we demonstrated that a correct treatment of CR
propagation can explain the occurrence of a decoupling region between
gas and magnetic field that in turn affects the disc formation.
We emphasise, however, that our calculations are not strictly self-consistent, because 
we computed microscopic resistivities using the density and magnetic field strength obtained
from {\em ideal} MHD simulations. If diffusion processes were included self-consistently in the numerical
simulation itself, the line twisting would be presumably reduced. This would attenuate the
effect of cosmic-ray mirroring, and the decrease in \zhh\ would not be so strong as found in this paper.
In this sense, our study should be considered a proof of concept showing how
a correct evaluation of \zhh\ can affect the protostellar disc formation.

\acknowledgements
MP and PH acknowledge the financial support of the Agence National
pour la Recherche (ANR) through the COSMIS project. 
MP and DG also acknowledge the support of the CNRS-INAF PICS project ``Pulsar wind nebulae,
supernova remnants and the origin of cosmic rays''.
This work has been carried out thanks to the support of the OCEVU Labex (ANR-11-LABX-0060) 
and the A*MIDEX project (ANR-11-IDEX-0001-02) funded by the ``Investissements d'Avenir'' French government programme managed by the ANR. 

\appendix

\section{Chemical model}
\label{app:chemcode}

In this Appendix we describe the ``minimal'' chemical model introduced 
in Sect.~\ref{ionfraction} to compute the ionisation fraction. The charged species considered
are $\rm H^+$, $\rm H_3^+$, molecular
ions $m$H$^+$ (e.g. HCO$^+$), ``metal'' ions $M^+$ (e.g. Mg$^+$), electrons $e$, and 
negatively charged dust grains, whereas the neutral species are H$_2$, heavy molecules $m$ 
(e.g. CO), metal atoms $M$, (e.g. Mg) and neutral grains.  Charged and neutral grains are collectively 
indicated as $g$. We indicate with $x(i)$ the abundance of each species $i$ with respect to H$_2$. 
The abundance of the neutral species is fixed. In particular, we assume $x(m)
\simeq 6\times10^{-4}$ and $x(M)\simeq 4\times10^{-8}$. All
rate coefficients are estimated at $T=10$~K.  

\subsection{Minimal chemical network}
Protons are produced
by CR ionisation of H$_2$ at a rate $\epsilon\zeta^{\rm
H_{2}}$, with $\epsilon\simeq0.05$ (Shah \& Gilbody~\cite{sg82}).
They are mainly destroyed by charge transfer (CT) with molecules
(at a rate $\beta\simeq10^{-9}$~cm$^{3}$~s$^{-1}$) and by recombination
on grains (at a rate $\alpha_{\rm gr}$, see Eq.~\ref{agr})
\be
\label{nhp}
\epsilon\zeta^{\rm H_{2}}n({\rm H_{2}})=
[\beta n(m{\rm H^{+}}) + \alpha_{\rm gr}n(g)]n({\rm H^{+}})\,.
\ee
The formation of $\rm H_{3}^{+}$ is driven by CR ionisation of $\rm
H_{2}$ at a rate $(1-\epsilon)\zeta^{\rm H_{2}}$, while destruction
is due to CT with heavy molecules, dissociative recombination (DR,
at a rate $\alpha_{\rm dr}\simeq10^{-6}$~cm$^{3}$~s$^{-1}$), and
recombination on grains
\be
\label{nhhhp}
(1-\epsilon)\zeta^{\rm H_{2}}n({\rm H_{2}})=
[\beta n(m{\rm H^{+}}) + \alpha_{\rm dr}n(e) +
\alpha_{\rm gr}n(g)]n({\rm H_{3}^{+}})\,.
\ee
The formation of molecular ions $m$H$^+$ occurs by CT of ${\rm
H}_3^+$ and heavy molecules, while destruction occurs by DR and
recombination on grains
\be
\label{nmhp}
\beta n({\rm H}_3^+)n(m{\rm H}^+) = 
[\alpha_{\rm dr}n(e) + \alpha_{\rm gr}n(g)]n(m{\rm H}^+)\,.
\ee
Metal ions are formed by CT with $\rm H_{3}^{+}$ and
$m{\rm H^{+}}$, and destroyed by recombination with free electrons and
on grains
\be
\label{nmp}
\beta n(m^{+})[n({\rm H_{3}^{+}})+n(m{\rm H^{+}})]=
[\alpha_{\rm rec}n(e)+\alpha_{\rm gr}n(g)]n(m^{+}).
\ee
Note that CT with metal atoms can be neglected with respect to DR
if $x(e)\gg(\beta/\alpha_{\rm dr})x(m^{+})\simeq10^{-3}x(m^{+})$.

Dust grains are assumed to be negatively charged (charge $-1$) or neutral.
The total number density of grains is obtained from the MRN size
distribution (Mathis et al.~\cite{mrn77})
\be
\label{MRN}
\frac{\ud n(g)}{\ud a}=Ca^{-3.5}\,,
\ee
between a minimum ($a_{\rm min}$) and a maximum ($a_{\rm max}$)
grain radius (see also Sect.~\ref{grainsize}). The normalisation
constant $C$ is obtained by imposing that the mass density of grains
is equal to a fraction $q=0.01$ of the gas density. We assume that
grains are spherical and have density $\rho_{g}=2$~g~cm$^{-3}$
(Flower et al.~\cite{fp05}).  For $a_{\rm max}\gg a_{\rm min}$ we
obtain
\be
C = \frac{3qm_{\rm H}}{4\pi\rho_{g}a_{\rm max}^{0.5}}n(\rm H_{2}).
\ee
Under these assumptions, the number density of grains is
\be\label{ngr}
n(g)=\frac{3qm_{\rm H}}{10\pi\rho_{g}a_{\rm max}^{0.5}a_{\rm min}^{2.5}}
n(\rm H_{2})\,,
\ee
and is strongly dependent on $a_{\rm min}$.

The coefficient of recombination of positive ions on negatively
charged grains was computed by Draine \& Sutin~(\cite{ds87}) assuming
the MRN size distribution,
\begin{eqnarray}\label{agr}
\alpha_{\rm gr}&=&\frac{10}{3}e^{2}a_{\rm min}
\left(\frac{8\pi}{m_{\rm i}kT}\right)^{1/2}
\left[1+\frac{3kTa_{\rm min}}{2e^{2}}(1-\psi)\right]\\\nonumber
&=&1.6\times10^{-7}\left(\frac{a_{\rm min}}{\AA}\right)
\left(\frac{T}{10~{\rm K}}\right)^{-1/2}\\\nonumber
&&\times\left[1+3.6\times10^{-4}\left(\frac{a_{\rm min}}{\AA}\right)
\left(\frac{T}{10~{\rm K}}\right)\left(\frac{1-\psi}{4}\right)\right]\,,
\end{eqnarray}
where $\psi$ is a numerical coefficient equal to $-2.5$ for an $e$--H$^+$
plasma and $-3.8$ for a heavy-ion plasma.
For simplicity, we adopt the same value of $\alpha_{\rm gr}$ for
all positively charged ions, assuming a typical ion mass $m_{\rm
i}=25m_{\rm H}$.  The actual value of $\alpha_{\rm gr}$ is larger
by a factor 5 and 3 for H$^+$ and H$_3^+$, respectively.

\subsection{Evaluation of the ionisation fractions}

The equation of charge neutrality is
\be
n(g)+n(e)=n({\rm H}^+)+n({\rm H}_3^+)+n(m{\rm H}^+)+n(m^+)\,.
\ee
Equations~(\ref{nhp}), (\ref{nhhhp}), (\ref{nmhp}), and (\ref{nmp}) then become
\be
x({\rm H}^+)=\frac{\epsilon A}{q(m)+rx(g)}\,,
\ee
\be
x({\rm H}_3^+)=\frac{(1-\epsilon) A}{q(m)+rx(g)+x(e)}\,,
\ee
\be
x(m{\rm H}^+)=\frac{(1-\epsilon) A}
{[rx(g)+x(e)][q(m)+rx(g)+x(e)]}q(m)\,,
\ee
\be
x(M^+)=\frac{(1-\epsilon) A}{[rx(g)+sx(e)][rx(g)+x(e)]}q(M)\,,
\ee
where we have defined 
\be
A=\frac{\zeta^{{\rm H}_2}}{n({\rm H}_2)\alpha_{\rm dr}}, \;
q(m)=\frac{\beta x(m)}{\alpha_{\rm dr}},\;
q(M)=\frac{\beta x(M)}{\alpha_{\rm dr}}, \;
r=\frac{\alpha_{\rm gr}}{\alpha_{\rm dr}}, \;
s=\frac{\alpha_{\rm rec}}{\alpha_{\rm dr}}.
\ee
Thus, the equation of charge neutrality 
\be
x(g)+x(e)=x({\rm H}^+)+x({\rm H}_3^+)+x(m{\rm H}^+)+x(m^+),
\ee
becomes
\be
x(g)+x(e)=\frac{\epsilon A}{q(m{\rm H}^+)+rx(g)}+
\frac{(1-\epsilon) A}{rx(g)+x(e)} 
+\frac{(1-\epsilon) Aq(m^+)}{[rx(g)+sx(e)][rx(g)+x(e)]}.
\label{neutreq}
\ee
We solve Eq.~(\ref{neutreq}) as a cubic equation for $x(e)$ assuming
that all grains have charge $-1$. When $x(e)$ becomes negative, we
set $x(e)=0$ and we solve Eq.~(\ref{neutreq}) for $x(g)$.

\section{Dependence of diffusion coefficients on grain radius}
\label{app:meangrain}

\begin{figure}[]
\begin{center}
\includegraphics[width=.45\textwidth]{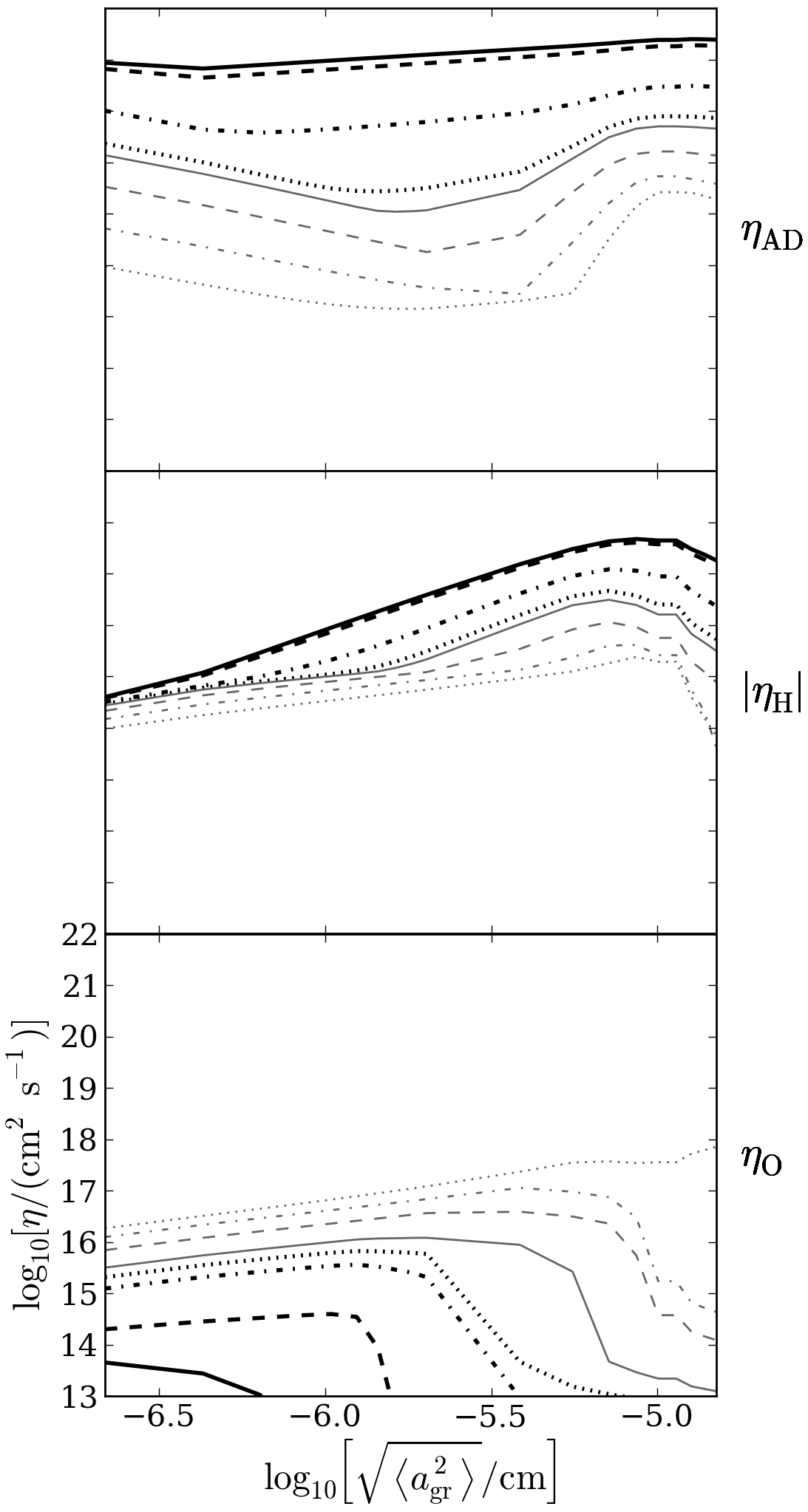}
\caption[]{Ambipolar ({\em upper panel}), Hall ({\em middle panel}), and Ohmic ({\em lower panel}) diffusion
coefficients as a function of the mean grain radius 
computed at different molecular hydrogen densities: $10^{5}$ ({\em thick solid line}), 
$10^{6}$ ({\em thick dashed line}),
$10^{7}$ ({\em thick dash-dotted line}),
$10^{8}$ ({\em dotted line}),
$10^{9}$ ({\em thin solid line}),
$10^{10}$ ({\em thin dashed line}),
$10^{11}$~cm$^{-3}$ ({\em thin dash-dotted line}),
and $10^{12}$~cm$^{-3}$ ({\em thin dotted line}).}
\label{eta_vs_amin}
\end{center}
\end{figure}

In this Appendix we compute the resistivities
using Eqs.~(\ref{eqetaAD})--(\ref{eqetaO}) varying the minimum
radius of the grain size distribution and fixing the H$_2$ density
to evaluate the sensitivity of the diffusive terms
to the grain size. For the magnetic field strength we assume 
$|{\bf B}|=(n/{\rm cm^{-3}})^{0.47}$~$\mu$G (Crutcher~\cite{c99}), in order
to be independent on specific models.
The MRN size distribution given by
Eq.~(\ref{MRN}) implies that the mean value of the square of the
grain radius weighted on the grain distribution (namely the quantity
that enters the equation for the momentum transfer rate coefficients),
defined by
\be 
\langle a_{\rm gr}^{2}\rangle=\frac{\int_{a_{\rm min}}^{a_{\rm
max}}a^{2}Ca^{-3.5}\ud a}{\int_{a_{\rm min}}^{a_{\rm max}}Ca^{-3.5}\ud
a}=5\frac{a_{\rm max}^{-0.5}-a_{\rm min}^{-0.5}}{a_{\rm max}^{-2.5}-a_{\rm
min}^{-2.5}}\,, 
\ee 
is close to $a_{\rm min}^{2}$.  We vary the minimum grain radius
between $10^{-7}$ to $10^{-5}$~cm, fixing $a_{\rm max}=3\times10^{5}$~cm.
This corresponds to values of $\langle a_{\rm gr}^{2}\rangle^{1/2}$
ranging from $2.2\times10^{-7}$ to $1.5\times10^{-5}$~cm.
Figure~\ref{eta_vs_amin} shows that the coefficient of ambipolar diffusion 
is not monotonic with the grain radius but presents an
absolute minimum at larger radii with increasing H$_2$ densities.
On the other hand, the Hall term increases with the grain radius for any
value of $n({\rm H}_2)$, starting to decrease only at very large
grain size ($a_{\rm min}\gtrsim10^{-5}$~cm).  Finally, the Ohmic
resistivity becomes important at high densities
($n\gtrsim10^{11}$~cm$^{-3}$) and increases monotonically with grain
size.

\section{Contributions to the diffusion time of the magnetic field}
\label{app:sepcontrdifftime}

\begin{figure}[]
\begin{center}
\includegraphics[width=0.5\textwidth]{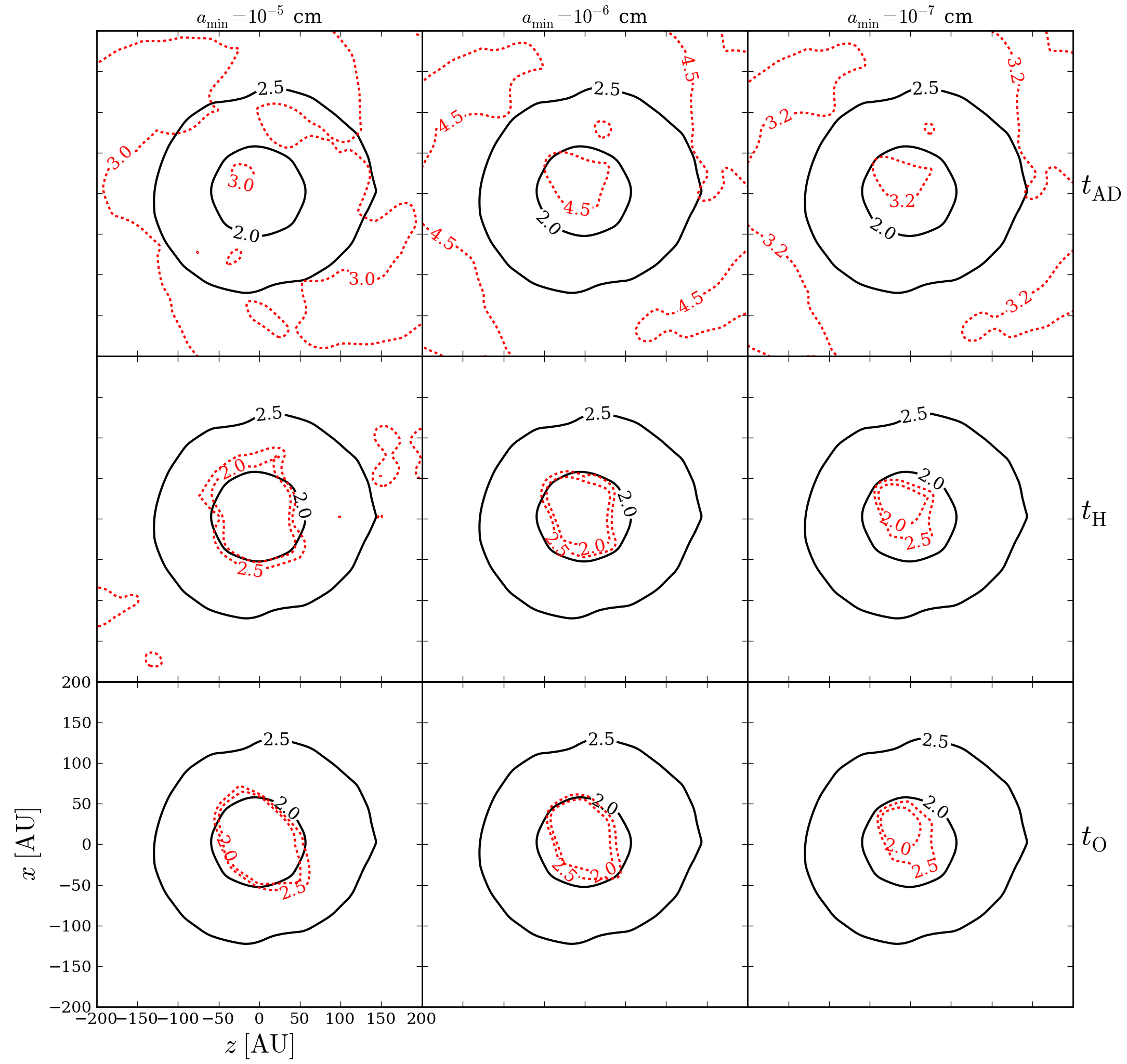}
\caption[]{Ambipolar, Hall, and Ohmic contribution to the diffusion
time ({\em red dotted lines}) compared with the dynamical time ({\em
black solid lines}).  Labels show $\log_{10}(t/{\rm yr})$.}
\label{drifttime_ADHO_plotpaper_ZX_187}
\end{center}
\end{figure}

For the sake of completeness, in Fig.~\ref{drifttime_ADHO_plotpaper_ZX_187}
we show separately the diffusion time for ambipolar, Hall, and
Ohmic diffusion computed with \zhh\ from model $L_2$.  As expected,
at the high densities reached in the disc region, ambipolar diffusion
time is more than one order of magnitude larger than
the dynamical time scale. On the contrary, Hall and Ohmic diffusion
times are comparable and lower than the dynamical time scale in a region
whose radius decreases from about 50~AU to 25~AU with
decreasing minimum grain size. This is another way to say that
gas-magnetic field decoupling is due to Hall and Ohmic
diffusion at densities higher than $n\sim10^{10}$~cm$^{-3}$.

\begin{thebibliography}{100}

\bibitem[2000]{awt00}
Andr\'e, P., Ward-Thompson, D. \& Barsony, M. 2000, Protostars and
Planets IV, 59
%
\bibitem[2002]{ba02}
Belloche, A., Andr\'e, P., Despois, D. \& Blinder, S. 2002, \aap, 393, 927
%
\bibitem[2012a]{bw12a}
Braiding, C.~R. \& Wardle, M. 2012a, MNRAS, 422, 261
%
\bibitem[2012b]{bw12b}
Braiding, C.~R. \& Wardle, M. 2012b, MNRAS, 427, 3188
%
\bibitem[1995]{b95}
Bodenheimer, P. 1995, ARA\&A, 33, 199
%
\bibitem[1978]{cv78}
Cesarsky, C.~J. \& V\"olk, H.~J. 1978, \aap, 70, 367
%
\bibitem[1994]{cm94}
Ciolek, G.~E; \& Mouschovias, T.~Ch. 1994, \apj, 425, 142
%
\bibitem[1995]{cm95}
Ciolek, G.~E; \& Mouschovias, T.~Ch. 1995, \apj, 454, 194
%
\bibitem[2013]{cabv13} Cleeves, L.~I., Adams, 
F.~C., Bergin, E.~A., \& Visser, R.\ 2013, \apj, 777, 28 
%
\bibitem[2011]{chh11}
Commer\c con, B., Hennebelle, P. \& Henning, T. 2011, \aap, 530, 13
%
\bibitem[1999]{c99}
Crutcher, R. M. 1999, \apj, 520, 706
%
\bibitem[2012]{c12}
Crutcher, R.~M. 2012, ARA\&A, 50, 29 
%
\bibitem[2010]{db10}
Dapp, W.~B. \& Basu, S. 2010, \aap, 521 256
%
\bibitem[1987]{ds87}
Draine, B.~T. \& Sutin, B. 1987, \apj, 320, 803
%
\bibitem[2005]{fp05}
Flower, D.~R., Pineau des For\^ets, G. \& Walmsley, C.~M.
2005, \aap, 436, 933
%
\bibitem[2006]{fh06}
Fromang, S., Hennebelle, P. \& Teyssier, R. 2006, \aap, 457, 371
%
\bibitem[2006]{gl06}
Galli, D., Lizano, S., Shu, F.~H. \& Allen, A.
2006, \apj, 647, 374
%
\bibitem[2008]{hf08}
Hennebelle, P. \& Fromang, S. 2008, \aap, 477, 9
%
\bibitem[2009]{hc09}
Hennebelle, P. \& Ciardi, A. 2009, \aap, 506, L29
%
\bibitem[2011]{hc11}
Hennebelle, P., Commer\c con, B., Joos, M., Klessen, R.~S., Krumholz, M. et al.
2011, \aap, 528, A72
%
\bibitem[2012]{jh12}
Joos, M., Hennebelle, P. \& Ciardi, A. 2012 \aap, 543, 128
%
\bibitem[2013]{jh13}
Joos, M., Hennebelle, P., Ciardi, A. \& Fromang, S. 2013 \aap, 554, A17
%
\bibitem[2011]{kl11}
Krasnopolsky, R., Li, Z.-Y. \& Shang, H. 2011 \apj, 733, 54
%
\bibitem[2012]{kl12}
Krasnopolsky, R., Li, Z.-Y. \& Shang, H. et al. 2012 \apj, 757, 77
%
\bibitem[2011]{lc11}
Labadens, M., Chapon, D., Pomar\'ede, D. \& Teyssier, R.
2011, ADASS XXI Proceedings, 461, 837
%
\bibitem[2011]{mi11}
Machida, M.~N., Inutsuka, S.-I. \& Matsumoto, T. 2011, PASJ, 63, 555
%
\bibitem[1977]{mrn77}
Mathis, J.~S., Rumpl, W. \& Nordsieck, K.~H. 1977, \apj, 217, 425
%
\bibitem[1989]{m89}
McKee, C.~F. 1989, \apj, 345, 782
%
\bibitem[2008]{ml08}
Mellon, R.~R. \& Li, Z.-H. 2008, \apj, 681, 1356
%
\bibitem[2009]{ml09}
Mellon, R.~R. \& Li, Z.-H. 2009, \apj, 698, 922
%
\bibitem[1956]{ms56}
Mestel, L. \& Spitzer, L., Jr. 1963, MNRAS, 116, 503
%
\bibitem[2013]{mu13}
Murillo, N.~M., Lai, S.-P., Bruderer, S., Harsono, D., \& van Dishoeck, E.~F.\ 2013, \aap, 560, A103 
%
\bibitem[2002]{nn02}
Nakano, T., Nishi, R. \& Umebayashi, T.
2002, ApJ, 573, 199
%
\bibitem[2009]{pgg09}
Padovani, M., Galli, D. \& Glassgold, A.~E. 2009, \aap, 501, 619 (PGG09)
\bibitem[2011]{pg11}
Padovani, M. \& Galli, D. 2011, \aap, 530, A109 (PG11)
%
\bibitem[2013]{phg13}
Padovani, M., Hennebelle, P. \& Galli, D. 2013, \aap, 560, A114 (PHG13)
%
\bibitem[2008]{pgb08}
Pinto, C., Galli, D. \& Bacciotti, F.
2008, \aap, 484, 1
%
\bibitem[2008]{pg08}
Pinto, C. \& Galli, D.
2008, \aap, 484, 17
%
\bibitem[2013]{sd13}
Santos-Lima, R., de Gouveia Dal Pino, E.~M. \& Lazarian, A. 2013, MNRAS, 429, 3371
%
\bibitem[2012]{sb12}
Seifried, D., Banerjee, R., Pudritz, R.~E. et al. 2012, MNRAS, 423, 40
%
\bibitem[1982]{sg82}
Shah, M.~B. \& Gilbody, H.~B. 1982, J. Phys. B, 15, 3441
%
\bibitem[2006]{sg06}
Shu, F.~H., Galli, D., Lizano, S. \& Cai, M.
2006, \apj, 647, 382
%
\bibitem[1968]{st68}
Spitzer, L. \& Tomasko, M. G. 1968, \apj, 152, 971
%
\bibitem[2000]{sm00} 
Strong, A.~W., Moskalenko, I.~V., Reimer, O.\ 2000, \apj, 537, 763 
%
\bibitem[2012]{ts12}
Takakuwa, S., Saito, M., Lim, J., Saigo, K., Sridharan, T.~K. et al. 2012, \apj, 754, 52
%
\bibitem[2002]{t02}
Teyssier, R. 2002, \aap, 385, 337
%
\bibitem[2012]{th12}
Tobin, J.~J., Hartmann, L., Chiang, H.-F., Wilner, D.~J., Looney, L.~W. et al. 2012, Nature, 492, 83
%
\bibitem[1981]{un81} 
Umebayashi, T. \& Nakano, T. 1981, PASJ, 33, 617
%
\bibitem[2009]{un09}
Umebayashi, T. \& Nakano, T. 1990, \mnras, 243, 103
%
\bibitem[1999]{wn99}
Wardle, M. \& Ng, C. 1999, MNRAS, 303, 239
%
\bibitem[2007]{w07}
Wardle, M. 2007, Ap\&SS, 311, 35
%
\bibitem[1998]{w98} 
Webber, W.~R.\ 1998, \apj, 506, 329 
%
\bibitem[2011]{wc11}
Williams, J.~P. \& Cieza, L.~A. 2011, ARA\&A, 49, 67

\end{thebibliography}
\end{document}